# TMSR: Tiny Multi-path CNNs for Super Resolution

Chia-Hung Liu
*Department of Computer Science and Information Engineering*
*National Cheng Kung University*
Tainan, Taiwan
chweb.liu@msa.hinet.net

Tzu-Hsin Hsieh
*Department of Computer Science and Information Engineering*
*National Cheng Kung University*
Tainan, Taiwan
celinehsieh68@gmail.com

Kuan-Yu Huang
*Department of Computer Science and Information Engineering*
*National Cheng Kung University*
Tainan, Taiwan
p78091523@gs.ncku.edu.tw

Pei-Yin Chen
*Department of Computer Science and Information Engineering*
*National Cheng Kung University*
Tainan, Taiwan
pychen@mail.ncku.edu.tw

*Abstract*—In this paper, we proposed a tiny multi-path CNN-based Super-Resolution (SR) method, called TMSR. We mainly refer to some tiny CNN-based SR methods, under 5k parameters. The main contribution of the proposed method is the improved multi-path learning and self-defined activated function. The experimental results show that TMSR obtains competitive image quality (i.e. PSNR and SSIM) compared to the related works under 5k parameters.

## I. INTRODUCTION

Research on Single Image Super-Resolution (SISR) continues to thrive in both the academic and industrial sectors. With technological advancements, the size of photosensitive elements has evolved from an initial 100×100 grayscale pixels to the high-pixel standards of today over the course of 48 years. Therefore, the need arises to enlarge our earliest stored low-resolution image data to fit current high-resolution displays, which is the focus of current SISR research.

Well-known algorithms include bicubic interpolation, bilinear interpolation, and nearest-neighbor interpolation, all of which have proven sufficient for most image-scaling applications thus far. However, when enlarging these images to higher resolutions, undesirable artifacts such as aliasing occur. To address the issue of roughness in enlarged images, the SRCNN[1] (Super-Resolution Convolutional Neural Network) was proposed by Dong Chao and others from the University of Hong Kong in 2014. This Artificial Intelligence (AI) application enhances existing pixel data in images through deep learning up-sampling methods[4] and generates reasonable high-resolution image data, thereby reducing aliasing effects and improving image quality.

While entering the 21st century, the exponential growth in big data and computational speed has led to breakthrough developments in AI. Techniques for reconstructing high-resolution images from low-resolution ones using AI have far surpassed the effects achieved through interpolation methods. However, literature and empirical observations show that deeper and more complex neural network models, although significantly improving imaging results, come at the cost of longer training times and increased computational expenses. Nowadays, with the ubiquity of streaming platforms and digital imaging, designing digital display products like TVs, digital cameras, portable gaming devices, and our most commonly used smartphones must account for bandwidth limitations and hardware costs. Therefore, the focus of this paper is to explore how to build neural network models with fewer parameters that can still achieve better imaging results, allowing users to experience clear images across different hardware platforms.

## II. RELATED WORKS

Recently, Deep Neural Networks (DNNs), particularly Convolutional Neural Networks (CNNs), have proven to exhibit exceptional performance in various fields of computer vision. They are commonly used for classification, object detection, image segmentation, and other image-related problems. In this paper, we will introduce how CNNs are applied to Single Image Super-Resolution (SISR), contributing to the resolution of a variety of other problems associated with computer vision. Several CNN-based Super-Resolution (SR) methods have been proposed [1]-[7]. These CNN architectures consist of multiple layers of convolution and nonlinear functions, aiming to generate high-quality High-Resolution (HR) images in SR models. Initially, most research on Single Image Ultra-High Resolution in AI models focused on enhancing the model's prediction capabilities. By "prediction," we mean that after batch-learning training, these ultra-high-resolution models can enlarge new or untrained image data by various factors, while still maintaining excellent performance in terms of Peak Signal-to-Noise Ratio (PSNR).

However, in this paper, we must find a balance between speed and cost. Therefore, we propose a cost-effective super-resolution method based on CNN. We employ features of FSRCNN [2] for (i) no pre-processing of the input image, (ii) Non-linear mapping, and combine it with MobileNets[5] for (iii) Depthwise separable convolutions, and (iv) 1x1 convolutions to reduce the training dimensions. We also adopt VDSR [3] residual connection techniques. For residual connections, we use distributed multi-layer residual links to reduce the number of neuron parameters used in the entire neural network model.

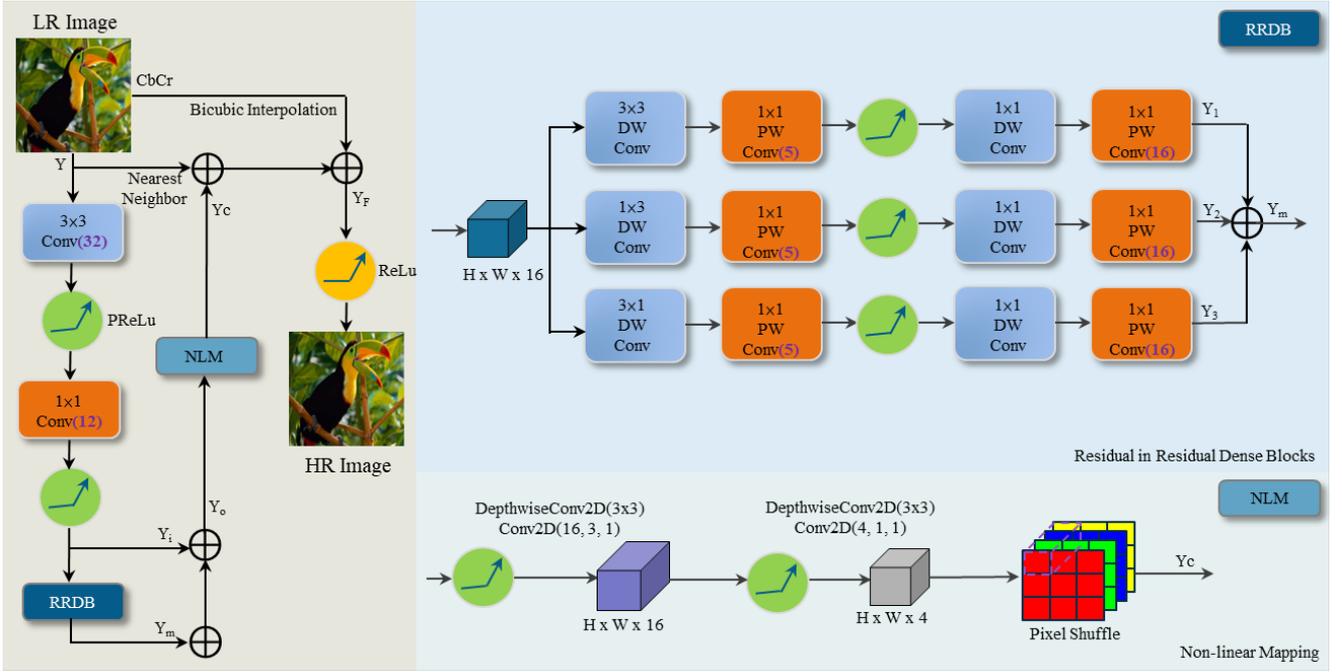

Figure 1. A block diagram of our proposed TMSR CNN-based SR network.

*A. Reduce the quantity of parameters in the CNN filter*

**Multi-path residual connection.**

One of the challenges in training neural networks is that we often require deeper and more layered networks to achieve better accuracy and performance. However, the deeper the network, the more difficult it is to converge during training. In our proposed Two-Stage Multi-Scale Residual (TMSR) model, we will employ two-stage residual connections, a simple yet highly effective technique that makes the training of deep neural networks more manageable. In traditional neural networks, data operates by sequentially forwarding the output of each layer to the input of the next layer. In contrast, residual connections skip some layers, providing an alternative pathway for the data to reach the latter part of the neural network, as shown in Figure 2.

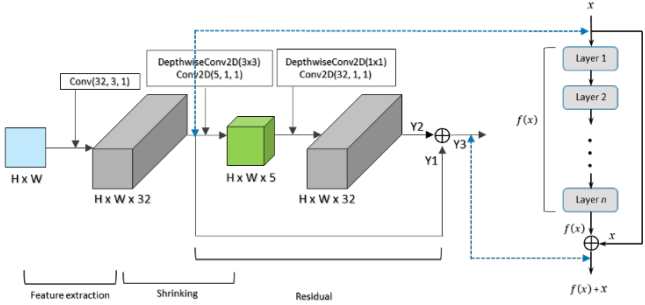

Figure 2.

In our proposed Super-Resolution (SR) model, we define the residual image $Y_2 = Y_3 - Y_1$, as shown in the blue-boxed area of Figure 2. Most of its values are likely zero or very small. After feature extraction to obtain $Y_1$, we compress the extracted feature information and immediately restore these features to their original dimension sizes, obtaining the restored $Y_2$ values for residual connections. Furthermore, our experiments reveal that adding multiple pathways for convolutional operations in the Residual Block layer before performing residual connections can improve the PSNR and SSIM values of the reconstructed image, as illustrated in the architecture in Figure 1.

In pursuit of cost-effectiveness and for comparison with existing SR models, we have tried to adjust the filter sizes in the model's Residual Block layer to approach a neuron parameter count of under 2.5K. Our approach is inspired by ResNeXt[6], presented at CVPR 2017, as shown in Figure 3. It manages to improve accuracy without increasing the number of parameters, simplifies the model architecture, and modularizes it. Most importantly, it reduces the

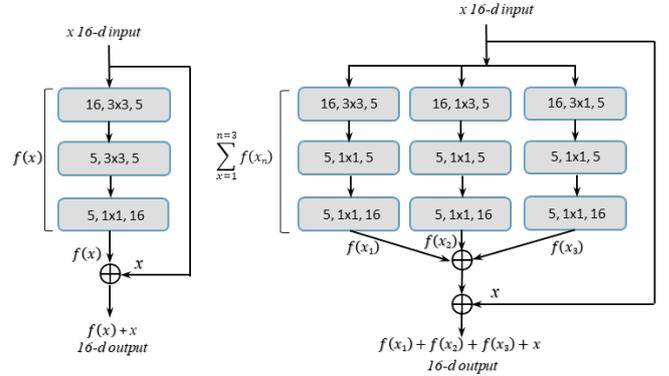

Figure 3. **Left**: A block of ResNet[6]. **Right**: A block of ResNetX[6] with cardinality = 3, with roughly the same complexity. A layer is shown as (input channels, filter size, output channels)

number of parameters

In our Two-Stage Multi-Scale Residual (TMSR) model, we have appropriately modified the original architecture of ResNeXt[6]. Similar to ResNeXt, we also start by dividing the original high-dimensional convolutional layer into multiple convolutional layers. However, while ResNeXt groups convolutional layers of the same dimensions, in our TMSR model, we modify the size of the convolutional kernels, adopting 3×3, 1×3, and 3×1 different kernel sizes. Then, convolutional operations are performed, and finally, these convolutional layers are fused together using residual connections. The term "cardinality," mentioned in the ResNeXt paper, refers to the number of divisions or groups. In our proposed model, the cardinality value is 3.

### B. Receptive field

**Vertical and Horizontal of Receptive field.**

Since we have already split the single-path residual connections into three layers using the ResNeXt[6] architecture, as mentioned earlier, we have also modified the original ResNeXt[6] structure. Through experimental observation, we found that if all three grouped pathways use a 3×3 convolutional kernel size for convolutional operations, the final output image quality is not better than when using different kernel sizes. Therefore, we have changed the convolutional operations in each layer, which initially used the same kernel size, to utilize different sizes.

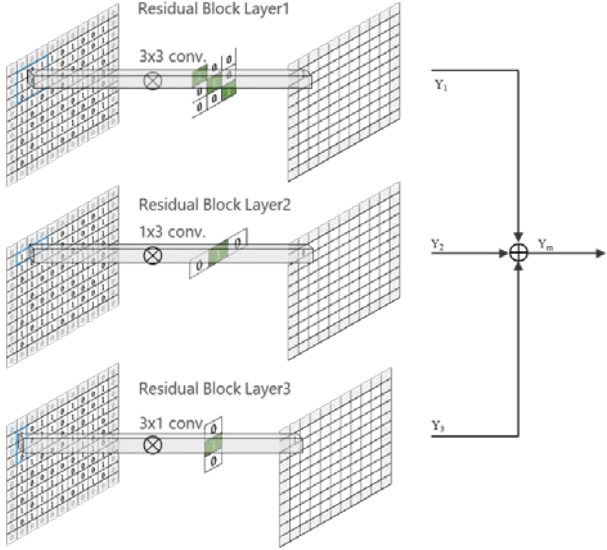

Figure 4. Vertical and Horizonital of Receptive field

By modifying the kernel sizes, we discovered that in multi-path residuals, adding convolutional kernels in both horizontal and vertical directions not only expands the receptive field but also yields higher PSNR values after residual connection operations. This is because the network structure can better utilize the information between the context in the images and the model, as shown in Figure 4. The kernel sizes for the three-path residual connections are 3×3, 1×3, and 3×1, respectively. We use these different sizes and orientations of receptive fields to verify their impact on image reconstruction.

### C. Activation Function

Another point worth noting is that our TMSR model employs the PReLU activation function instead of the standard ReLU. The ReLU activation function is non-zero-centered and is non-differentiable at zero, as shown in Figure 5(a). However, it is differentiable everywhere else. Another issue we observe in ReLU is the Dying ReLU problem, where some ReLU neurons essentially become inactive, remaining non-responsive regardless of the type of input provided. If there is no gradient flow and a large number of dead neurons exist in the neural network, its performance will be adversely affected. Therefore, we use Parametric ReLU, which, unlike the fixed slope of 0.01 used in Leaky ReLU, as shown in Figure 5(b), adjusts the parameter a according to the model for x < 0 as per Equation (2-13). By utilizing weights and biases, we can adjust the parameters learned through backpropagation across multiple layers.

$$PReLU(x) = \begin{cases} x, & if\ x > 0 \\ ax, & otherwise \end{cases} \quad (2\text{-}1)$$

### III. PROPOSED METHOD

#### A. Implementation Detail

**Training dataset.**

For objectivity and fairness, we utilize the T91 image dataset mentioned in the existing literature FSRCNN[2] as the sample set for model training. The T91 dataset is widely used as a training set for learning-based Super-Resolution (SR) methods. While deep models usually benefit from large datasets, our research found that a mere 91 images are insufficient for the deep model to achieve optimal performance. To make full use of the image dataset, we adopt data augmentation[9] to increase the number of images, providing the model with more learning references. We expand the dataset in two ways: (i) We scale all 91 images in the T91 dataset by factors of 0.9, 0.8, 0.7, and 0.6, and (ii) we further rotate the scaled images by 90, 180, and 270 degrees. After data augmentation, we will have $5 \times 4 - 1 = 19$ times the original number of images, resulting in a total training dataset of $5 \times 4 \times 91 = 1820$ images, as shown in Equation 3-1.

$$(1820 - 91(original\ image)) \div 19 = 91 \quad (3\text{-}1)$$

**Test and validation dataset.**

To ensure the fairness of the experiments, we utilized the most widely-used test datasets, such as Set5 and Set14. Additionally, we employed the B100 dataset, which consists of natural images from the Berkeley segmentation dataset used for benchmark testing by Timofte et al. and Yang & Yang. Lastly, we also used the Urban100 dataset, a recently provided city image dataset by Huang et al. The Urban100 dataset is particularly interesting as it contains many challenging images where existing methods fail.

**Training samples.**

We generated 1,820 images through data augmentation and are now prepared to commence the feature extraction for training data. First, we extract $f_{sub} \times f_{sub}$ sub-images from the original High-Resolution (HR) training images with a stride k of 14 pixels. The extracted $f_{sub} \times f_{sub}$ sub-images have dimensions of 32 pixels in both height and width. The extraction begins by moving horizontally across the HR image with a stride of 14 pixels to the right. Upon reaching the width boundary, the process returns to the starting point and moves vertically downward, again with a stride of 14 pixels, until the limits of both the image width and height are reached. This feature extraction process results in paired HR/LR images within the width and height constraints of each of the 1,820 images. Since the dimensions of each image vary, we eventually obtained 240,288 sub-images. We are aware that having more training samples can prevent overfitting during the training process. Additionally, our experiments found that if we also perform feature extraction with augmented samples from the **General 100** training dataset, we would obtain 3,820 images. With additional augmentation on the 3,820 extracted feature images from **General 100**, we would have 1,218,292 sub-images. Although the training time for our TMSR model is extended, the quality of the enlarged reconstructed images improves by 0.02dB~0.03dB.

**Training strategy.**

We use the Mean Square Error (MSE), as defined by Equation 3-2, as the loss function for training the images to enhance their reconstruction. We then calculate the Peak Signal-to-Noise Ratio (PSNR), as outlined in Equation 3-3, to evaluate the effectiveness of the reconstructed images. A higher PSNR value indicates better reconstruction quality. We also assess the excellence of the neural network model by calculating the PSNR values.

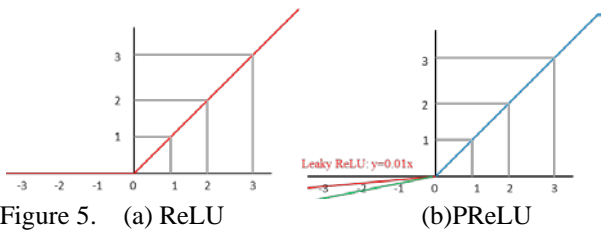

Figure 5.   (a) ReLU            (b)PReLU

| Methods | Parm. | Data Sets | | | | | | | | | | | |
|---|---|---|---|---|---|---|---|---|---|---|---|---|---|
| | | Set5 | | | Set14 | | | B100 | | | Urban100 | | |
| | | PSNR | SSIM | Time(s) | PSNR | SSIM | Time(s) | PSNR | SSIM | Time(s) | PSNR | SSIM | Time(s) |
| Bicubic | - | 33.66 | 0.9296 | - | 30.34 | 0.869 | - | 29.54 | 0.8434 | - | 26.88 | 0.841 | - |
| SRCNN | 8K | 36.34 | 0.9521 | 0.0596 | 32.18 | 0.9039 | 0.0933 | 31.11 | 0.8835 | - | 29.09 | 0.8897 | - |
| FSRCNNs | 4K | 36.57 | 0.9531 | 0.0511 | 32.28 | 0.9049 | 0.057 | 31.23 | 0.8866 | 0.0542 | 29.23 | 0.8914 | 0.0599 |
| HED-SR | 2.58K | 36.49 | 0.9538 | - | 32.29 | 0.9053 | - | 31.18 | 0.8862 | - | 29 | 0.889 | - |
| RTSRCNN | 2.56K | 36.56 | 0.9531 | 0.0448 | 32.43 | 0.9097 | 0.051 | 31.26 | 0.8867 | 0.049 | 29.24 | 0.9001 | 0.0564 |
| TMSR | **2.474K** | **36.66** | **0.9538** | 0.0484 | **32.49** | **0.9104** | 0.0589 | **31.32** | **0.888** | 0.0525 | **29.33** | **0.9012** | 0.0586 |

Table 1. Performance Comparisons Of Various SR Methods

$$L(\Theta) = \frac{1}{n}\sum_{i=1}^{n}\|F(\mathbf{Y}_i;\Theta) - \mathbf{X}_i\|^2 \quad (3\text{-}2)$$

$$PSNR = 10 \times log_{10}\frac{MAX_I^2}{MSE} \quad (3\text{-}3)$$

Additionally, we calculate the Structural Similarity Index Measure (SSIM), as defined in Equation 3-4. The SSIM index is computed over different images. The metric between two co-sized images x and y, both of size N×N, is evaluated as follows:

$$SSIM(x,y) = \frac{(2\mu_x\mu_y + C_1)(2\sigma_{xy} + C_2)}{(\mu_x^2 + \mu_y^2 + C_1)(\sigma_x^2 + \sigma_y^2 + C_2)} \quad (3\text{-}4)$$

Where $\mu_x$ is the mean of x, $\mu_y$ is the mean of y, $\sigma_x^2$ is the variance of x, and $\sigma_y^2$ is the variance of y. A SSIM value closer to 1 implies that the SR model is more successful in reconstructing images that are similar to the High-Resolution (HR) images in terms of brightness, contrast, and structure. For initialization, the filter weights for each layer start from zero. We design the convolutional filters using PReLU as the activation function. The training cycle is set for 5000 epochs, and the learning rate is set at 0.0001. Through experimentation, we found that a smaller learning rate greatly aids in the convergence of the network.

*B. Different Configuration*

We experimented by changing the activation function in our proposed TMSR model from PReLU to ReLU. As can be observed from Table 2, there is a decrease in the PSNR values by 0.02~0.06dB in the Set5 and Set14 validation datasets. This suggests that using PReLU as the activation function after the convolution operation helps to increase the PSNR value, thereby implying that the TMSR model produces higher-quality reconstructed output images.

| Methods | Data Sets | | | |
|---|---|---|---|---|
| | Set5 | | Set14 | |
| | PSNR | SSIM | PSNR | SSIM |
| TMSR w/ ReLu | 36.6 | 0.9534 | 32.47 | 0.9009 |
| TMSR w/ PReLu | 36.66 | 0.9538 | 32.49 | 0.9104 |

Table 2. Comparison of PSNR and SSIM Values Between ReLU and PReLU

Although using PReLU as the activation function after the convolution operation incurs a computational cost, we observe from Table 3. that utilizing PReLU as the activation function takes an additional 10 milliseconds compared to using ReLU.

| Methods | Data Sets | | | |
|---|---|---|---|---|
| | Set5 | Set14 | B100 | Urban100 |
| | Time(s) | Time(s) | Time(s) | Time(s) |
| TMSR w/ ReLu | 0.0406 | 0.0533 | 0.0458 | 0.05 |
| TMSR w/ PReLu | 0.0484 | 0.0612 | 0.0543 | 0.0586 |

Table 3. Comparison of Time Spent for Output Using ReLU and PReLU

## IV. EXPERIMENT RESULTS

We employ the TMSR model here to validate against Set5, Set14, B100, and Urban100 datasets. We calculate and compare the quantitative image quality differences using Equations (3-3) and (3-4). Table 1. summarizes the data for all the validation datasets. We observe that the TMSR model outperforms other CNN-based SR models in terms of both PSNR and SSIM values. Additionally, in Table IV, we list the average PSNR and SSIM values for the images in the Set5 dataset, while Table V presents the average PSNR and SSIM values for the images in the Set14 dataset.

| Set5 | upscaling factor | Bicubic | | TMSR | |
|---|---|---|---|---|---|
| | | - | | 2.474K | |
| images | Scale | PSNR | SSIM | PSNR | SSIM |
| baby | 2 | 37.05 | 0.95 | 38.37 | 0.96 |
| bird | 2 | 36.68 | 0.97 | 40.90 | 0.99 |
| butterfly | 2 | 27.45 | 0.91 | 32.78 | 0.97 |
| head | 2 | 34.85 | 0.86 | 35.67 | 0.88 |
| woman | 2 | 32.13 | 0.95 | 35.57 | 0.97 |
| Average | 2 | 33.63 | 0.9291 | **36.66** | 0.9538 |

Table 4. Set5 PSNR & SSIM value

| Set14 | upscaling factor | Bicubic | | TMSR | |
|---|---|---|---|---|---|
| | | - | | 2.474K | |
| images | Scale | PSNR | SSIM | PSNR | SSIM |
| baboon | 2 | 24.65 | 0.70 | 25.56 | 0.77 |
| barbara | 2 | 27.93 | 0.84 | 28.47 | 0.87 |
| coastguard | 2 | 29.13 | 0.79 | 30.82 | 0.85 |
| comic | 2 | 26.05 | 0.85 | 28.68 | 0.92 |
| face | 2 | 34.83 | 0.86 | 35.65 | 0.88 |
| flowers | 2 | 30.42 | 0.90 | 33.34 | 0.94 |
| foreman | 2 | 32.61 | 0.95 | 34.47 | 0.97 |
| lenna | 2 | 34.71 | 0.91 | 36.61 | 0.93 |
| man | 2 | 29.26 | 0.85 | 31.01 | 0.89 |
| monarch | 2 | 32.95 | 0.96 | 37.66 | 0.98 |
| pepper | 2 | 33.00 | 0.91 | 35.18 | 0.92 |
| ppt3 | 2 | 27.13 | 0.95 | 31.14 | 0.98 |
| zebra | 2 | 30.72 | 0.91 | 33.74 | 0.94 |
| Average | 2 | 30.26 | 0.8744 | 32.49 | 0.9104 |

Table 5. Set14 PSNR & SSIM value

Figure 6 displays the original HR images along with the effects of HR image reconstruction using bicubic interpolation and CNN-based SR methods, which include our proposed TMSR and RTSRCNN[7]. Despite our method using the least number of parameters, the reconstructed HR images still maintain sharp edges and minimal artifacts. Furthermore, in Figure 10, the HR images outputted by our proposed TMSR model in the challenging Urban100 dataset visually demonstrate superior image quality compared to other CNN-based SR methods.

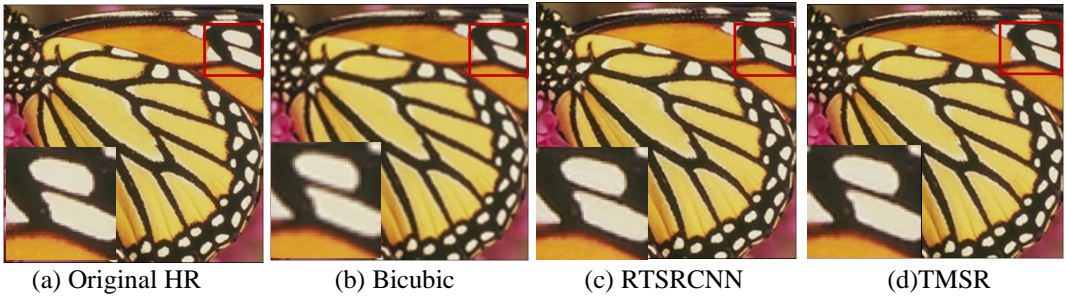

(a) Original HR   (b) Bicubic   (c) RTSRCNN   (d)TMSR

Figure 6. Comparison of HR output images (×2 on Monarch). (a) Original HR image. (b) Bicubic. (c) RTSRCNN[7]. (d) TMSR

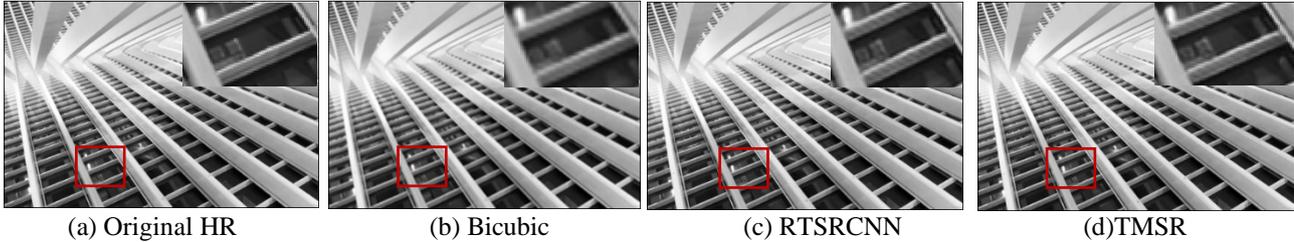

(a) Original HR   (b) Bicubic   (c) RTSRCNN   (d)TMSR

Figure 7. Comparison of HR output images (×2 on Urban100-042). (a) Original HR image. (b) Bicubic. (c) RTSRCNN[7]. (d) TMSR

## V. CONCLUSION

Upon observing current deep learning-based SR models, it's evident that deeper and more complex network architectures are utilized to maintain high-quality output results following image enlargement. However, the trade-off is the increased need for more neural network parameters to achieve these outputs. To address this, we have synthesized the best features of FSRCNN[2] and RTSRCNN[7] architectures, optimizing and redesigning a new SR model that not only maintains a certain level of image output quality but also uses 40% fewer neural parameters than FSRCNN and 4% fewer than RTSRCNN.

Through extensive experimental data, we have demonstrated that our proposed TMSR model yields satisfactory results. Due to the model's relatively small number of parameters, it can also be implemented on hardware platforms. In comparison to SR models like VDSR[3] and FSRCNN[2], the PSNR values of the output reconstruction on the Set5 test dataset reach 37.53dB and 37.0dB, respectively. This leaves a minor gap in comparison to our proposed SR model.

In the future, to close this gap, we could consider improving the model. Beyond hyperparameter tuning, our experiments also indicate that multi-path residual connections effectively boost image output quality. Alternatively, collecting or augmenting more training samples during the training phase could enhance the quality of reconstructed images by leveraging additional training features.

## REFERENCES

[1] C. Dong, C. C. Loy, K. He, and X. Tang, ''Learning a deep convolutional network for image super-resolution,'' in Proc. Eur. Conf. Comput. Vis. (ECCV), Cham, Switzerland: Springer, pp. 184–199, 2014.. *(references)*

[2] C. Dong, C. C. Loy, and X. Tang, ''Accelerating the super-resolution convolutional neural network,'' in Proc. Eur. Conf. Comput. Vis. (ECCV), Cham, Switzerland: Springer, pp. 391–407, 2016.

[3] Jiwon Kim, Jung Kwon Lee, Kyoung Mu Lee, ''Accurate Image Super-Resolution Using Very Deep Convolutional Networks,'' Proceedings of the IEEE Conference on Computer Vision and Pattern Recognition (CVPR), pp. 1646-1654, 2016.

[4] SHI, Wenzhe, et al, "Real-time single image and video super-resolution using an efficient sub-pixel convolutional neural network," In: Proceedings of the IEEE conference on computer vision and pattern recognition, pp. 1874-1883, 2016.

[5] HOWARD, Andrew G., et al. "Mobilenets: Efficient convolutional neural networks for mobile vision applications," in arXiv preprint arXiv:1704.04861, 2017.

[6] XIE, Saining, et al. Aggregated residual transformations for deep neural networks. In: Proceedings of the IEEE conference on computer vision and pattern recognition. 2017. p. 1492-1500.

[7] Y. Kim, J.-S. Choi, and M. Kim, ''A real-time convolutional neural network for super-resolution on FPGA with applications to 4K UHD 60 fps video services,'' in IEEE Trans. Circuits Syst. Video Technol., vol. 29, no. 8, pp. 2522–2524, Aug. 2019.

[8] LEE, Sumin, et al, "CNN acceleration with hardware-efficient dataflow for super-resolution" in IEEE Access, pp. 187754-187765, Aug. 2020.

[9] YOO, Jaejun; AHN, Namhyuk; SOHN, Kyung-Ah, "Rethinking data augmentation for image super-resolution: A comprehensive analysis and a new strategy," In: Proceedings of the IEEE/CVF Conference on Computer Vision and Pattern Recognition, pp. 8375-8384, 2020.

[10] AYAZOGLU, Mustafa, "Extremely lightweight quantization robust real-time single-image super resolution for mobile devices," In: Proceedings of the IEEE/CVF Conference on Computer Vision and Pattern Recognition, pp. 2472-2479, 2021.